\documentclass[a4paper,12pt]{article}
\usepackage{amsthm,amssymb,amsmath,amsfonts,mathrsfs,url,hyperref}

\title{Thermal Equilibrium Distribution in Infinite-Dimensional Hilbert Spaces}
\author{
Roderich Tumulka\footnote{Fachbereich Mathematik,
	Eberhard-Karls-Universit\"at, 
	Auf der Morgenstelle 10, 72076 T\"ubingen, Germany.
     E-mail: roderich.tumulka@uni-tuebingen.de}
}
\date{November 19, 2019}

\theoremstyle{plain}
\newtheorem{thm}{Theorem}
\newtheorem{lemma}{Lemma}

\newtheorem{definition}{Definition}

\newcommand{\ket}[1]{\vert#1\rangle}
\newcommand{\bra}[1]{\langle#1\vert}
\newcommand{\Tr}{\mathrm{tr}}
\newcommand{\CCC}{\mathbb{C}}
\newcommand{\RRR}{\mathbb{R}}
\newcommand{\NNN}{\mathbb{N}}
\newcommand{\EEE}{\mathbb{E}}

\renewcommand{\Re}{\mathrm{Re}}
\renewcommand{\Im}{\mathrm{Im}}
\newcommand{\scp}[2]{\langle #1| #2 \rangle}
\newcommand{\Hilbert}{\mathscr{H}}
\newcommand{\sphere}{\mathbb{S}}
\newcommand{\C}{\mathscr{C}}
\newcommand{\Borel}{\mathscr{B}}
\newcommand{\D}{\mathscr{D}}
\newcommand{\G}[1]{G(#1)}
\newcommand{\GA}[1]{GA(#1)}
\newcommand{\GAP}[1]{GAP(#1)}
\newcommand{\be}{\begin{equation}}
\newcommand{\ee}{\end{equation}}

\begin{document}
\maketitle
\begin{abstract}
The thermal equilibrium distribution over quantum-mechanical wave functions is a so-called Gaussian adjusted projected (GAP) measure, $GAP(\rho_\beta)$, for a thermal density operator $\rho_\beta$ at inverse temperature $\beta$. More generally, $GAP(\rho)$ is a probability measure  on the unit sphere in Hilbert space for any density operator $\rho$ (i.e., a positive operator with trace 1). In this note, we collect the mathematical details concerning the rigorous definition of $GAP(\rho)$ in infinite-dimensional separable Hilbert spaces. Its existence and uniqueness follows from Prohorov's theorem on the existence and uniqueness of Gaussian measures in Hilbert spaces with given mean and covariance. We also give an alternative existence proof. Finally, we give a proof that $GAP(\rho)$ depends continuously on $\rho$ in the sense that convergence of $\rho$ in the trace norm implies weak convergence of $GAP(\rho)$.

\bigskip

\noindent Key words: 
Gaussian measure, GAP measure, Scrooge measure,
canonical ensemble in quantum mechanics.
\end{abstract}

\section{Introduction}

The Gaussian adjusted projected (GAP) measures 
are certain probability distributions over quantum-mechanical wave functions
that arise as the typical distribution of the conditional wave function 
for a quantum system in thermal equilibrium with a heat
bath at a given temperature \cite{Gold1,GLMTZ15}.
The goal of this note is to provide a rigorous foundation of the measure $GAP(\rho)$ for every density operator $\rho$ in the case of an infinite-dimensional separable Hilbert space $\Hilbert$. After we formulate the precise definition, the existence and uniqueness of $GAP(\rho)$ will follow from corresponding theorems about Gaussian measures on such Hilbert spaces. We then provide an explicit construction (and thereby an alternative existence proof) of $GAP(\rho)$ and prove the continuous dependence of $GAP(\rho)$ on $\rho$.

The family of GAP measures is a family of probability measures on
Hilbert spaces, first considered in \cite{JRW94} under the name ``Scrooge measure'' as the most spread-out measure on the unit sphere with a given density matrix, then in \cite{Gold1} as playing the role of a
thermal equilibrium distribution of the wave function of a quantum
system. 
The GAP measures relevant to thermal equilibrium are those
associated with the canonical density operators
\begin{equation}\label{rhobeta}
  \rho_\beta = \tfrac{1}{Z} e^{-\beta H}\,,
\end{equation}
where $Z= \Tr \, e^{-\beta H}$ is the normalization constant,
$\beta$ the inverse temperature and $H$ the Hamiltonian. A detailed
discussion of GAP measures and their physical background can be found 
in \cite{Gold1,Rei08,GLMTZ15}. See \cite{PT13} for a discussion of conditional wave functions with spin in thermal equilibrium, and \cite{TZ05} for a rigorous study about the support of GAP measures (that took existence and uniqueness for granted), in particular
about what $GAP(\rho_\beta)$-typical wave functions look like.

\section{Mean, Covariance, and Density Operator}

We begin by introducing some notation. Let $\Hilbert$ be a
separable complex Hilbert space, and for any topological space $X$
let $\Borel(X)$ denote its Borel $\sigma$-algebra.
We denote the unit sphere in $\Hilbert$ by
\begin{equation}
  \sphere(\Hilbert) = \{\psi \in \Hilbert: \|\psi\|=1\}\,.
\end{equation}
Note that $\sphere(\Hilbert)$ lies in $\Borel(\Hilbert)$ (because it is the pre-image of $\{1\}$ under the continuous function $\|\cdot\|$), and that a subset of $\sphere(\Hilbert)$ lies in $\Borel(\sphere(\Hilbert))$ if and only if it lies in $\Borel(\Hilbert)$ (because the open sets in $\sphere(\Hilbert)$ are the intersections of open sets in $\Hilbert$ with $\sphere(\Hilbert)$). In particular, a measure on $\bigl(\sphere(\Hilbert),\Borel(\sphere(\Hilbert))\bigr)$ can be regarded as a measure on $(\Hilbert,\Borel(\Hilbert))$ that is concentrated on $\sphere(\Hilbert)$.

\begin{definition}
Let $\mu$ be a probability measure on $(\Hilbert,\Borel(\Hilbert))$. The vector $\psi_0\in\Hilbert$ is called the \textbf{mean} of $\mu$ if and only if, for every $\phi\in\Hilbert$,
\be\label{meandef}
\scp{\phi}{\psi_0} = \int \mu(d\psi) \, \scp{\phi}{\psi}\,.
\ee
The operator $C_\mu:\Hilbert\to\Hilbert$ is called the \textbf{covariance operator} of $\mu$ if and only if, for every $\phi,\chi\in\Hilbert$,
\be\label{covdef}
\scp{\phi}{C_\mu\,\chi} = \int\mu(d\psi) \, \scp{\phi}{\psi-\psi_0}\scp{\psi-\psi_0}{\chi}\,.
\ee
\end{definition}

The mean of $\mu$ need not exist (e.g., the function $\psi\mapsto \scp{\phi}{\psi}$ may fail to be $\mu$-integrable), but if it exists it is unique. (Indeed, if $\scp{\phi}{\psi_0'} = \scp{\phi}{\psi_0}$ for every $\phi\in\Hilbert$ then $\psi_0'=\psi_0$.) The definition of the covariance operator applies only to those $\mu$ that have a mean. Even in that case, the covariance operator may not exist, but if it exists it is unique and positive. (Indeed, an operator $C_\mu$ is uniquely determined by the sesqui-linear form $(\phi,\chi)\mapsto \scp{\phi}{C_\mu\,\chi}$; positivity is immediate from \eqref{covdef}.) 

Let $\D(\Hilbert)$ be the set of all density operators,
i.e., of all positive operators in the trace class with trace 1. 

\begin{lemma}\label{lem:rho}
Every probability measure $\mu$ on $\bigl(\sphere(\Hilbert),\Borel(\sphere(\Hilbert))\bigr)$ possesses a mean and a covariance, as well as 
a unique operator $\rho_\mu:\Hilbert\to\Hilbert$ such that
\begin{equation}\label{rhodef}
\scp{\phi}{\rho_\mu\,\chi}=\int \mu(d\psi)\, \scp{\phi}{\psi}\scp{\psi}{\chi}
\end{equation}
for all $\phi,\chi \in \Hilbert$; $\rho_\mu$, called the density operator of $\mu$, lies in $\D(\Hilbert)$. If the mean of $\mu$ is zero then the density operator coincides with the covariance operator.
\end{lemma}

\proof
We first consider the mean. For any $\phi\in\Hilbert$, the function
\be
f:\sphere(\Hilbert)\to\CCC\,,\quad
f(\psi)=\scp{\phi}{\psi}
\ee
is continuous and thus measurable; it is also bounded by $\|\phi\|$ and thus, in particular, $\mu$-integrable; therefore,
\be
L(\phi):=\int \mu(d\psi)\,f(\psi)
\ee
is a bounded conjugate-linear form on $\Hilbert$ with $\|L\|\leq 1$. By the Riesz lemma, there is a unique $\psi_0\in\Hilbert$ such that $L(\phi) = \scp{\phi}{\psi_0}$, which is equivalent to \eqref{meandef}. 

We now turn to the density operator.
For any $\phi,\chi\in\Hilbert$, the function
\be\label{fdef}
f:\sphere(\Hilbert)\to\CCC\,, \quad
f(\psi) = \scp{\phi}{\psi} \scp{\psi}{\chi}
\ee
is measurable and bounded,
\be\label{fbound}
|f(\psi)| \leq \|\phi\| \, \|\chi\|\,.
\ee
Therefore, $f$ is $\mu$-integrable, and 
\be
B(\phi,\chi) := \int \mu(d\psi) \, f(\psi)
\ee
is a bounded sesqui-linear form on $\Hilbert$ with $\|B\|\leq 1$. By standard arguments using the Riesz lemma, there is a unique operator $\rho_\mu:\Hilbert\to\Hilbert$ such that $B(\phi,\chi)=\scp{\phi}{\rho_\mu\, \chi}$, which is equivalent to \eqref{rhodef}. 

Since, for $\phi=\chi$, $f(\psi)$ is always real and non-negative, $\rho_\mu$ is positive (and, in particular, self-adjoint). In order to compute $\Tr\, \rho_\mu$, choose any orthonormal basis $\{\varphi_n:n\in\NNN\}$ of $\Hilbert$ and use the monotone convergence theorem to interchange summation and integration in
\begin{align}
\Tr \, \rho_\mu 
&= \sum_n \scp{\varphi_n}{\rho_\mu\, \varphi_n}
= \sum_n \int \mu(d\psi)\, \bigl| \scp{\varphi_n}{\psi} \bigr|^2\\
&= \int \mu(d\psi)\sum_n \bigl| \scp{\varphi_n}{\psi} \bigr|^2
=\int\mu(d\psi) \, \|\psi\|^2 
= 1\,.
\end{align}

In order to prove the existence of the covariance operator, repeat the above argument for the density operator starting from \eqref{fdef} with $\psi$ replaced by $\psi-\psi_0$; \eqref{fbound} gets replaced by
\be
|f(\psi)| \leq \|\phi\| \,\bigl(1+\|\psi_0\|\bigr)^2 \, \|\chi\|\,,
\ee
which leads to $\|B\|\leq (1+\|\psi_0\|)^2$, and to a unique operator $C_\mu$ such that $B(\phi,\chi)=\scp{\phi}{C_\mu\, \chi}$, which is equivalent to \eqref{covdef}. 

The last statement of Lemma~\ref{lem:rho} is obvious.
\endproof

\section{Definition of the GAP Measure in Finite Dimension}

The measure $GAP(\rho)$ on (the Borel $\sigma$-algebra of)
$\sphere(\CCC^d)$, $d\in\NNN$, is built starting from the measure $G(\rho)$,
the Gaussian measure on $\CCC^d$ with mean zero and covariance matrix
$\rho$, which can be
defined as follows: Let $S$ be the subspace of $\CCC^d$ on which
$\rho$ is supported, i.e., its positive spectral subspace, or
equivalently the orthogonal complement of its kernel, or
equivalently its range; let $k=\dim S$ and $\rho_+$ the restriction
of $\rho$ to $S$; then $G(\rho)$ is a measure on $\CCC^d$
supported on $S$ with the following density relative to the Lebesgue
measure $\lambda$ on $S$:
\begin{equation}
  \frac{dG(\rho)}{d\lambda}(\psi) = \frac{1}{\pi^k \,
  \det \rho_+} \exp(-\langle \psi |\rho^{-1}_+| \psi \rangle)\,.
\end{equation}

Now we define the adjusted Gaussian measure $GA(\rho)$ on $\CCC^d$ as:
\begin{equation}
  GA(\rho)(d\psi) = \|\psi\|^2 G(\rho)(d\psi)\,.
\end{equation}
If $\Psi^{GA}$ is a $GA(\rho)$-distributed vector, then $GAP(\rho)$
is the distribution of the vector projected to the unit sphere, i.e., of
\begin{equation}
  \Psi^{GAP}:= \frac{\Psi^{GA}}{\|\Psi^{GA}\|}
\end{equation}
Like $\G{\rho}$, but not like $\GA{\rho}$, $\GAP{\rho}$ has covariance matrix $\rho$ (see \cite{Gold1} or Lemma~\ref{lem:cov} below).

\section{Definition of the GAP Measure in Infinite Dimension}

One can define for any measure $\mu$ on $\bigl(\Hilbert,\Borel(\Hilbert)\bigr)$ the ``adjust-and-project'' procedure. We denote by $A\mu$ the adjusted measure
\begin{equation}\label{Adef}
A\mu(d\psi) = \|\psi\|^2 \, \mu(d\psi)
\end{equation}
and the projection on the unit sphere is defined as:
\begin{equation}\label{Pdef}
P:\Hilbert \setminus \{0\} \to \sphere(\Hilbert)\, , \quad
P(\psi)= \frac{\psi}{\|\psi\|}\,.
\end{equation}
Then the adjusted-and-projected measure is $P_*  A\mu = A\mu \circ
P^{-1}$, where $P_*$ denotes the action of $P$ on measures. Since the function $P$ is measurable relative to $\Borel(\Hilbert)$ and $\Borel(\sphere(\Hilbert))$, $P_* A$ is a mapping from the measures on $\Hilbert$ to the measures on $\sphere(\Hilbert)$. If
\begin{equation}\label{expectation}
\int \mu(d\psi) \, \|\psi\|^2 = 1
\end{equation}
then the adjusted projected measure $P_*A\mu$ is a probability measure on $\sphere(\Hilbert)$.

\begin{lemma}\label{lem:cov}
Suppose $\mu$ is a probability measure on $\bigl( \Hilbert,\Borel(\Hilbert) \bigr)$ satisfying \eqref{expectation}. Then $\mu$ possesses a mean and a covariance operator. If, moreover, the mean of $\mu$ is zero, then the density operator of $P_*A\mu$ coincides with the covariance operator of $\mu$, $\rho_{P_*A\mu}=C_\mu$.
\end{lemma}

\proof
We first show that $\mu$ possesses a mean. From \eqref{expectation} we have that
\begin{align}
\int\mu(d\psi) \, \|\psi\| 
&= \biggl( \int_{\|\psi\|\leq 1} + \int_{\|\psi\|>1}\biggr) \mu(d\psi)\, \|\psi\|\\
&\leq \mu\{\psi\in\Hilbert: \|\psi\|\leq 1\} + \int_{\|\psi\|>1} \mu(d\psi) \, \|\psi\|^2\\
&\leq 1+\int\mu(d\psi) \, \|\psi\|^2 = 2\,.\label{intnormpsi}
\end{align}
The function
\be
f(\psi) := \scp{\phi}{\psi}\,,
\ee
which satisfies
\be
|f(\psi)| \leq \|\phi\| \, \|\psi\|\,,
\ee
is therefore $\mu$-integrable with
\be
\int \mu(d\psi) \, |f(\psi)| \leq 2\|\phi\|\,,
\ee
and now the same argument as in the proof of Lemma~\ref{lem:rho} proves the existence of $\psi_0$ satisfying \eqref{meandef}.

We now show that $\mu$ possesses a covariance operator. The function
\be
f(\psi):=\scp{\phi}{\psi-\psi_0}\scp{\psi-\psi_0}{\chi}\,,
\ee
which satisfies
\be
|f(\psi)| \leq \|\phi\| \, \bigl(\|\psi\|+\|\psi_0\|\bigr)^2\, \|\chi\|\,,
\ee
is by \eqref{expectation} and \eqref{intnormpsi} $\mu$-integrable with
\be
\int \mu(d\psi) \, |f(\psi)| \leq \|\phi\|\, \bigl(1+4\|\psi_0\| + \|\psi_0\|^2\bigr) \, \|\chi\| \,, 
\ee
and now the same argument as in the proof of Lemma~\ref{lem:rho} proves the existence of $C$ satisfying \eqref{covdef}.

Now assume that the mean of $\mu$ is zero.
For any $\phi\in\Hilbert$, by \eqref{rhodef},
\begin{align}
\scp{\phi}{\rho_{P_*A\mu}\,\phi} &=
\int_{\sphere(\Hilbert)} P_*A\mu(d\psi) \, \scp{\phi}{\psi} \scp{\psi}{\phi}\\
&=\int_{\Hilbert} A\mu(d\chi) \, \frac{\scp{\phi}{\chi}\scp{\chi}{\phi}}{\|\chi\|^2}\\
&=\int_{\Hilbert} \mu(d\chi) \, \scp{\phi}{\chi}\scp{\chi}{\phi}= \scp{\phi}{C_\mu\,\phi}\,.
\end{align}
The operator $\rho_{P_*A\mu}$ is uniquely determined by the quadratic form $\phi\mapsto \scp{\phi}{\rho_{P_*A\mu}\, \phi}$.
\endproof

We now turn to formulating the general definition of GAP measures.

\begin{definition}\label{def:GaussianRV}
The complex random variable $Z$ is \textbf{Gaussian with variance $\sigma^2$} if and only if $\Re\, Z$ and $\Im\, Z$ are independent real Gaussian random variables, both with variance $\sigma^2/2$. (We include delta measures among the Gaussian distributions, corresponding to $\sigma=0$.)
\end{definition}

Here, $\Re$ ($\Im$) denotes the real (imaginary) part. 
To put Definition~\ref{def:GaussianRV} differently, a complex Gaussian random variable is one whose distribution is either a delta measure concentrated in one point or has density 
\be
\frac{1}{\pi \sigma^2} \exp\biggl(-\frac{|z-z_0|^2}{\sigma^2}\biggr)
\ee
relative to the Lebesgue measure for some $z_0\in\CCC$ and $\sigma>0$. 

\begin{definition}\label{def:G}
A probability measure $\mu$ on
$(\Hilbert,\Borel(\Hilbert))$ is a \textbf{Gaussian measure} if and only if, for
every $\phi\in\Hilbert$, the random variable $\Hilbert \ni \omega
\mapsto \scp{\phi}{\omega} \in \CCC$ is Gaussian
when $\omega$ has distribution $\mu$.
\end{definition}

\begin{definition}
A probability measure $\nu$ on $\bigl(\sphere(\Hilbert),\Borel(\sphere(\Hilbert))\bigr)$ is a \textbf{GAP measure} if and only if $\nu =P_* A\mu$ for a suitable Gaussian measure $\mu$ as in Definition~\ref{def:G} with mean 0.
\end{definition}

\section{Existence and Uniqueness}

\begin{thm}\label{ex}
For every positive trace-class operator $\rho$ with $\Tr\,\rho=1$ on the separable Hilbert space $\Hilbert$ there exists a unique GAP measure with density operator $\rho$.
\end{thm}

We will infer this fact from the corresponding statement about Gaussian measures, due to Prohorov \cite{Pro56} (quoted from \cite[p.~29]{Kuo75}):

\begin{thm}\label{G} 
Every Gaussian measure $\mu$ on $\Hilbert$ possesses a mean $\psi_0\in\Hilbert$ and a covariance operator, which is a positive trace-class operator $C$ on $\Hilbert$. 
For every $\psi_0\in\Hilbert$ and every positive trace-class operator $C$ on $\Hilbert$, there is a unique Gaussian measure $\mu$ with mean $\psi_0$ and covariance $C$. 
\end{thm}

Prohorov proved this using characteristic functions; the characteristic function $\hat{\mu}$ of a measure $\mu$ is defined by 
\be\label{hatmudef}
\hat{\mu}(\psi) = \int \mu(d\phi) \,
\exp\bigl(i\Re \scp{\phi}{\psi} \bigr)\,.
\ee
In fact, the characteristic function of the Gaussian measure with mean $\psi_0$ and covariance operator $\rho$ is
\begin{equation}\label{hatmu}
  \hat\mu(\psi) = \exp\bigl(i\Re\scp{\psi_0}{\psi}-\scp{\psi}{\rho\, \psi} \bigr)\,.
\end{equation}

\bigskip

\proof[Proof of Theorem~\ref{ex}.]
We write $G(\rho)$ for the unique Gaussian measure with mean 0 and covariance $\rho$. Set, for any $\rho\in\D(\Hilbert)$, $\nu(\rho)=P_*AG(\rho)$. To see that $\nu(\rho)$ is a probability measure, we need to check \eqref{expectation} for $\mu=G(\rho)$. Indeed, using an orthonormal basis $\{\phi_n:n\in\NNN\}$ of $\Hilbert$, 
we can write
\be
\int G(\rho)(d\psi) \, \|\psi\|^2=
\int G(\rho)(d\psi) \, \sum_n |\scp{\phi_n}{\psi}|^2=
\ee
[by the monotone convergence theorem]
\be
=\sum_n \int G(\rho)(d\psi) \,  |\scp{\phi_n}{\psi}|^2=
\ee
[by \eqref{covdef}, since $G(\rho)$ has covariance $\rho$]
\be
=\sum_n  \scp{\phi_n}{\rho\, \phi_n}=\Tr\, \rho =1\,.
\ee

Now Lemma~\ref{lem:cov} yields that the density operator of $\nu(\rho)$ is $\rho$, which proves the existence part of Theorem~\ref{ex}. For the uniqueness part, any GAP measure $\nu$ with density operator $\rho$ is by definition of the form $\nu=P_*A\mu$ with $\mu$ a Gaussian probability measure with mean 0 on $(\Hilbert,\Borel(\Hilbert))$; by Lemma~\ref{lem:cov}, $\mu$ has density (or covariance) operator $\rho$, too; by Prohorov's theorem, $\mu=G(\rho)$; thus, $\nu=P_*AG(\rho)$.
\endproof

Instead of $\nu(\rho)$, we will from now on write $GAP(\rho)$.

\section{Explicit Construction (and Alternative Existence Proof)}

It is instructive to see an explicit construction of the measure $G(\rho)$, which by the way provides an alternative proof of the existence statement in Theorem~\ref{G}. To this end, we need the following lemma.

\begin{lemma}\label{CsubB}
Let $\CCC^\NNN$ be the vector space of all complex sequences, let
$\C$ be the $\sigma$-algebra on $\CCC^\NNN$ generated by the cylinder sets, let $\ell^2\subset \CCC^\NNN$ be the space of square-summable sequences, and let $\Borel(\ell^2)$ be the Borel $\sigma$-algebra arising from the Hilbert space norm in $\ell^2$. Then $\ell^2\in\C$ and $\Borel(\ell^2) \subset \C$. Moreover, the cylinder sets in $\ell^2$ (i.e., the sets of the form $T\cap \ell^2$, where $T$ is a cylinder set in $\CCC^\NNN$) generate $\Borel(\ell^2)$. 
\end{lemma}

\proof[Proof of Lemma~\ref{CsubB}.] 
For every $N\in\NNN$ and $K>0$, the set
\be
S_{N,K}:=\bigl\{(x_n)_{n\in\NNN}\in\CCC^\NNN: |x_1|^2+\ldots+|x_N|^2<K\bigr\}
\ee
is a cylinder set; thus 
\be
\ell^2 = \bigcup_{K\in\NNN}\bigcap_{N\in\NNN}S_{N,K} \in\C\,.
\ee

Note that every cylinder set in $\ell^2$ lies in $\Borel(\ell^2)\cap \C$. It is a known theorem that the Borel $\sigma$-algebra of a Polish space (and a separable
Hilbert space is a Polish space) has the property that any countable
family of Borel sets that separates points generates the full
$\sigma$-algebra \cite[Thm.~3.3.5]{Arv76}. 
There is clearly a countable family $\mathscr{T}:=\{T_n:n\in\NNN\}$ of cylinder sets in $\ell^2$ that separates any two points in $\ell^2$; so, $\mathscr{T}$ generates $\Borel(\ell^2)$. Since $T_n\in\C$, the $\sigma$-algebra generated by $\mathscr{T}$ is contained in $\C$. As a further consequence, the family of all cylinder sets in $\ell^2$ generates $\Borel(\ell^2)$.
\endproof

We now construct, for every positive trace-class operator $C$, a Gaussian measure $\mu$ with mean 0 and covariance $C$. Translation by any vector $\psi_0\in\Hilbert$ will then provide a Gaussian measure with mean $\psi_0$ and covariance $C$. Since trace-class operators are compact, $C$ possesses an orthonormal basis $\{\varphi_n:n\in\NNN\}$ of eigenvectors; let $p_n$ be the eigenvalue of $\varphi_n$; that is,
\be
C=\sum p_n \ket{\varphi_n}\bra{\varphi_n}\,.
\ee
Let $Z_n$ be independent complex Gaussian random variables with mean 0 and variances $p_n$. By the Kolmogorov extension theorem \cite{Bill}, there exists such a random sequence $(Z_n)$ in $\CCC^\NNN$, i.e., there exists the appropriate product measure $\tilde{\mu}$ on the $\sigma$-algebra $\C$ generated by the cylinder sets. Lemma~\ref{CsubB} ensures that $\tilde{\mu}$ defines a measure on $\Borel(\ell^2)$, $\mu'=\tilde{\mu}|_{\Borel(\ell^2)}$. The random sequence $(Z_n)$ is almost surely square-summable because the expectation of its $\ell^2$-norm $\sum_n |Z_n|^2$ is finite,
\begin{equation}\label{Esumvariances}
  \EEE \sum_n |Z_n|^2 = \sum_n \EEE |Z_n|^2 = \sum_n p_n = \Tr\, C,
\end{equation}
and therefore $\sum_n|Z_n|^2$ is almost surely finite, $\tilde{\mu}(\ell^2)=1$, so $\mu'$ is a probability measure. The unitary isomorphism $\ell^2\to\Hilbert$ defined by the orthonormal basis $\{\varphi_n\}$ translates $\mu'$ into a measure $\mu$ on $(\Hilbert,\Borel(\Hilbert))$, which is the distribution of the random vector
\begin{equation}
\Psi^G := \sum_n Z_n \ket{n}\,.
\end{equation}
The measure $\mu$ is Gaussian because for every $\psi\in\Hilbert$,
\begin{equation}
\scp{\psi}{\Psi^G} = \sum_n \scp{\psi}{n} Z_n\,,
\end{equation}
which is a limit of linear combinations of complex Gaussian random variables and thus has a Gaussian distribution in $\CCC$. Furthermore, $\mu$ has mean 0 and covariance $C$.

\section{Continuous Dependence on $\rho$}

On $\D(\Hilbert)$ and
on the trace class of $\Hilbert$, the norm we use is the \emph{trace
norm}
\begin{equation}
  \|M\|_1 = \Tr |M| = \Tr \sqrt{M^* M}\,.
\end{equation}
We write $\mu_n \Rightarrow \mu$ to denote that the sequences of
measures $\mu_n$ (on a Hilbert space) \emph{converges weakly} to $\mu$.
This means that $\mu_n(f) \to \mu(f)$ for every bounded continuous
function $f$, where we have written
\begin{equation}\label{muf}
\mu(f):=\int \mu(d\psi) f(\psi)\,.
\end{equation}

\begin{thm}\label{contGAP}
Let $\Hilbert$ be a separable Hilbert space. The mapping $\rho\mapsto\GAP{\rho}$ from $\D(\Hilbert)$ to
the set of the probability measures on $\sphere(\Hilbert)$ is continuous in the sense that, for $\rho,\rho_n \in \D(\Hilbert)$ for every $n\in\NNN$:
\begin{equation}
\text{if }\|\rho_n -\rho\|_1 \to 0 
\text{ then }GAP(\rho_n) \Rightarrow GAP(\rho)\,.
\end{equation}
\end{thm}

The following lemma is the corresponding statement about
$G(\rho)$.

\begin{lemma}\label{contG}
If $\rho,\rho_n \in \D(\Hilbert)$ for every $n\in\NNN$ and $\|\rho_n
- \rho\|_1 \to 0$ then $G(\rho_n) \Rightarrow G(\rho)$.
\end{lemma}

\proof 
We use characteristic functions. As usual, the characteristic function $\hat{\mu}:\Hilbert\to\CCC$ of a probability measure $\mu$ on $\Hilbert$ is defined by \eqref{hatmudef}. 
We write $\mu_n = G(\rho_n)$ and
$\mu=G(\rho)$; their characteristic functions are:
\begin{equation}\label{hatGauss}
  \hat{\mu}_n(\psi) = \exp\bigl(-\bra{\psi}\rho_n\ket{\psi}\bigr)\,, \quad
  \hat{\mu}(\psi) = \exp\bigl(-\bra{\psi}\rho\ket{\psi}\bigr)\,.
\end{equation}

We use Lemma VI.2.1 on p.~153 of \cite{Par67}, which implies that
\textit{if a sequence $\mu_n$ of measures on $\Hilbert$ is
conditionally compact and the characteristic functions $\hat\mu_n$
converge pointwise to $\hat\mu$, then $\mu_n \Rightarrow \mu$.}
Here, $\mu = G(\rho)$ and $\mu_n = G(\rho_n)$.

The second condition, pointwise convergence of the characteristic
functions, follows from the convergence
of $\scp{\psi}{\rho_n\,\psi}$ to $\scp{\psi}{\rho\,\psi}$, which
follows from the convergence of $\rho_n$ to $\rho$ in the operator
norm, which follows from the convergence in the trace norm.

The other condition is conditional compactness. According to Theorem VI.2.2 on
p.~154 of \cite{Par67}, \emph{the sequence $\mu_n$ is conditionally
compact if
\begin{equation}
\sup_{n\in\NNN} \sum_{i=k}^\infty \scp{b_i}{\rho_n \, b_i} \to
0\quad \text{as }k\to\infty\,,
\end{equation}
where $\rho_n$ is the covariance operator of $\mu_n$ and $\{b_i\}$
an orthonormal basis of $\Hilbert$.} To see that this is the case,
fix any orthonormal basis $\{b_i\}$ and note that for the operator
$\Delta_n := |\rho_n-\rho|$ our hypothesis says $\Tr\, \Delta_n \to
0$ or, in other words, that for any $\varepsilon>0$ there is
$N\in\NNN$ such that, for all $n>N$,
\begin{equation}\label{largen}
  0\leq \Tr\, \Delta_n < \varepsilon\,.
\end{equation}
Since $\Tr\, \Delta_n < \infty$ for all $n\in\NNN$, there is
$K\in\NNN$ such that
\begin{equation}\label{smalln}
  \sum_{i=K}^\infty \scp{b_i}{\Delta_n \, b_i} < \varepsilon
  \quad\text{for all }n\leq N\,.
\end{equation}
Moreover,
\begin{equation}
  0\leq \sum_{i=k}^\infty \scp{b_i}{\rho_n \, b_i} \leq
  \sum_{i=k}^\infty \scp{b_i}{\rho \, b_i} + \sum_{i=k}^\infty \scp{b_i}{\Delta_n \, b_i}\,,
\end{equation}
where the first term on the right hand side is less than
$\varepsilon$ if $k$ is large enough because $\Tr\,\rho<\infty$, and
the second term is less than $\varepsilon$ for all $n\in\NNN$ if
$k>K$, either because of \eqref{smalln} (if $n\leq N$) or because of
\eqref{largen} (if $n>N$).
\endproof

We now establish the continuity of the ``adjustment'' mapping $A$
defined in \eqref{Adef}. 

\begin{lemma}\label{contA}
The mapping $A$ from the set of probability measures $\mu$ on
$\Hilbert$ such that $\int \mu(d\psi) \, \|\psi\|^2 = 1$ to the set of all probability measures on $\Hilbert$ is continuous. That is, suppose
that for every $n\in\NNN$, $\mu_n$ is a probability measure on
$(\Hilbert,\Borel(\Hilbert))$ such that $\int \mu_n(d\psi) \,
\|\psi\|^2 = 1$. If $\mu_n \Rightarrow \mu$ with $\int \mu(d\psi) \,
\|\psi\|^2 = 1$ then $A\mu_n\Rightarrow A\mu$.
\end{lemma}

\proof 
Fix $\varepsilon>0$ and an arbitrary bounded, continuous
function $f: \Hilbert \to \RRR$. Set $g(\psi) = \|\psi\|^2$. Since,
by hypothesis, $\mu(g) = 1$ in the notation \eqref{muf}, there
exists $R>0$ so large that
\begin{equation}
  \mu(1_{B_R} g) = \int_{B_R} \mu(d\psi) \, \|\psi\|^2 > 1- \frac{\varepsilon}{6\|f\|_\infty}\,,
\end{equation}
where $1_{B_R}$ denotes the indicator function of the set $B_R =
\{\psi\in \Hilbert: \|\psi\|<R\}$. Let the ``cut-off function''
$\chi: \Hilbert \to \RRR$ be any continuous function such that
$\chi=1$ on $B_R$, $\chi=0$ outside $B_{2R}$, and $0\leq \chi \leq
1$. Because $\chi g$ is a bounded continuous function, and because
$\mu_n \Rightarrow \mu$, we have that $\mu_n(\chi g) \to \mu(\chi
g)$, that is, there is an $n_1\in \NNN$ such that for all $n>n_1$,
\begin{equation}
  \left| \mu_n(\chi g) - \mu(\chi g) \right| < \frac{\varepsilon}{6\|f\|_\infty} \,.
\end{equation}
Therefore, using that $\chi \geq 1_{B_R}$,
\begin{equation}
  \mu_n(\chi g) > \mu(\chi g) - \frac{\varepsilon}{6\|f\|_\infty}
  \geq \mu(1_{B_R} g) - \frac{\varepsilon}{6\|f\|_\infty}
  > 1 - \frac{\varepsilon}{3\|f\|_\infty} \,,
\end{equation}
and thus that
\begin{equation}
  \Bigl|\mu_n\bigl( f(1-\chi)g \bigr)\Bigr| \leq \mu_n \bigl(| f(1-\chi)g |\bigr) \leq
  \|f\|_\infty \mu_n\bigl((1-\chi)g \bigr) =
\end{equation}
\begin{equation}
  = \|f\|_\infty \Bigl( \mu_n(g) - \mu_n(\chi g) \Bigr) <
  \|f\|_\infty \Bigl( 1- 1+ \frac{\varepsilon}{3\|f\|_\infty} \Bigr) = \varepsilon/3\,.
\end{equation}
Likewise,
\begin{equation}
 \Bigl|\mu\bigl( f(1-\chi) g \bigr) \Bigr| \leq \mu \bigl(| f(1-\chi)g |\bigr) \leq
  \|f\|_\infty \mu\bigl((1-\chi)g \bigr) \leq
\end{equation}
\begin{equation}
  \leq \|f\|_\infty \Bigl( \mu(g) - \mu(1_{B_R} g) \Bigr) <
  \|f\|_\infty \Bigl( 1- 1+ \frac{\varepsilon}{6\|f\|_\infty} \Bigr) = \varepsilon/6 <
  \varepsilon/3\,.
\end{equation}
Because $f\chi g$ is a bounded continuous function, and because
$\mu_n \Rightarrow \mu$, we have that $\mu_n(f\chi g) \to \mu(f\chi
g)$, that is, there is an $n_2\in\NNN$ such that for all $n>n_2$,
\begin{equation}
  \left| \mu_n(f\chi g) - \mu(f\chi g) \right| < \varepsilon/3\,.
\end{equation}
Thus,
\begin{equation}
  \left| A\mu_n(f) - A\mu(f) \right| = \left| \mu_n(fg) - \mu(fg) \right| \leq
\end{equation}
\begin{equation}
  \leq \left| \mu_n(f\chi g) - \mu(f\chi g) \right| +
  \left| \mu_n\bigl( f(1-\chi) g \bigr) \right| +
  \left| \mu\bigl( f(1-\chi)g \bigr) \right| < \varepsilon\,.
\end{equation}
\endproof

We remark that the hypothesis $\int \mu(d\psi) \|\psi\|^2=1$ cannot
be dropped, that is, does not follow from $\int \mu_n(d\psi)
\|\psi\|^2=1$. An example is $\mu_n = (1-1/n) \delta_0 + (1/n)
\delta_{\psi_n}$, where $\delta_\phi$ means the Dirac delta measure
at $\phi$ and $\psi_n$ is any vector with $\|\psi_n\|^2 = n$; then
$\mu_n$ is a probability measure with $\int \mu_n(d\psi)
\|\psi\|^2=1$ but $\mu_n \Rightarrow \delta_0$, which has $\int
\delta_0(d\psi) \|\psi\|^2=0$.

\bigskip

\proof[Proof of Theorem~\ref{contGAP}] 
Suppose $\|\rho_n-\rho\|_1\to 0$. We
have that $\GAP{\rho_n}=P_*A(\G{\rho_n})$, that $\int
\|\psi\|^2\G{\rho}=1$, and that $\left(AG(\rho)\right)(0)=0$. Since the projection
$\psi\mapsto P\psi$ to the unit sphere
as defined in \eqref{Pdef} is continuous (also in infinite dimension) at any $\psi\neq 0$, Theorem~\ref{contGAP}
follows from Lemma~\ref{contA} and Lemma~\ref{contG}.
\endproof

\bigskip

\noindent\textit{Acknowledgments.} 
I am grateful to Eric Carlen, Ernesto de Vito, Detlef D\"urr, Martin Kolb, Frank Loose, Enrico Massa, Ulrich Menne, Reiner Sch\"atzle, and Stefan Teufel for helpful discussions.

\end{document}